\documentclass[aps,pre,reprint,10pt,superscriptaddress,showpacs]{revtex4-1}
\usepackage{amsfonts} 
\usepackage{amsmath}
\usepackage{amssymb}
\usepackage{graphicx}%\[  \]
\usepackage{subfigure}
\usepackage{color}
\newcommand*\xbar[1]{%
  \hbox{%
    \vbox{%
      \hrule height 0.5pt % The actual bar
      \kern0.5ex%         % Distance between bar and symbol
      \hbox{%
        \kern-0.1em%      % Shortening on the left side
        \ensuremath{#1}%
        \kern-0.1em%      % Shortening on the right side
      }%
    }%
  }%
} 
\begin{document}
\title{Stimulated scattering instability in a  relativistic plasma}
\author{A. P. Misra}
\email{apmisra@visva-bharati.ac.in; apmisra@gmail.com}
\author{Debjani Chatterjee}
\email{chatterjee.debjani10@gmail.com}
\affiliation{Department of Mathematics, Siksha Bhavana, Visva-Bharati University, Santiniketan-731 235, West Bengal, India}
\pacs{52.25.Dg, 52.27.Ep, 52.35.Mw, 52.35.Sb}
\begin{abstract}
We study the stimulated scattering instabilities of an intense linearly polarized electromagnetic  wave (EMW)  in a relativistic plasma with degenerate electrons.  Starting from  a   relativistic hydrodynamic model and the Maxwell's equations, we derive  coupled nonlinear  equations for low-frequency electron and ion plasma oscillations that are driven by the EMW's ponderomotive force. The  nonlinear dispersion relations are then obtained from the coupled nonlinear equations which reveal stimulated Raman scattering (SRS), stimulated Brillouin scattering (SBS), and  modulational instabilities (MIs) of  EMWs. It is shown that    the thermal pressure of ions and  the relativistic degenerate pressure of electrons significantly modify the characteristics of SRS, SBS,  and MIs.
\end{abstract}
\maketitle
\section{Introduction} \label{sec-intro}
The nonlinear self-interactions of finite amplitude intense electromagnetic  waves (EMWs) and relativistic/nonrelativistic plasmas have received a significant research attention in recent years (see, e.g., Refs. \onlinecite{chanturia2017,shukla2012,singh2013,Salimullah1990,jain1984,rose1987,liu1996,walsh1984,gupta2015,parashar2013,gratton1997,gomberoff1997}). Such high-frequency (hf) EMWs are used for plasma heating, e.g., in inertial fusion plasmas \cite{kruer1973}, as well as for plasma diagnostics \cite{glenzer2009}, e.g., in solid density plasmas that are created by intense laser and charged particle beams. Furthermore, laser-plasma interaction provides a rich source of nonlinear phenomena including the formation of coherent structures as localized bursts of $x$ rays and  $\gamma$ rays \cite{begelman1984} from compact astrophysical objects, fast ignition, particle acceleration, generation of different kinds of waves and instabilities \cite{shukla2007}.
So, under certain conditions, the collective parametric  effects such as stimulated Raman and Brillouin scattering instabilities, and localization of high-frequency  (hf) EMWs could have a definite signature on the radiation spectra  (ranging from radio to $\gamma$ rays) of astrophysical objects \cite{zheleznyakov1996}.
On the other hand, for high-density plasmas, such as  as those in the interior of white dwarfs, neutron stars, and also at the source of $\gamma$-ray bursts  \cite{shapiro2004}, the relevant plasmas are relativistically degenerate and thus obey the Fermi-Dirac statistics.
\par
The nonlinear interaction of strong EMWs with electrostatic plasma oscillations has been considered by a number of authors. For example, Shukla and Stenflo \cite{shukla2006} studied  the stimulated scattering instabilities in an ultracold quantum plasma. Stenflo and Brodin \cite{stenflo2010} considered the effects of quantum particle dispersion (associated with the Bohm potential) to advance the theory of large amplitude circularly polarized  EMWs in a quantum plasma.    In a recent study, it has been shown that the stimulated Raman scattering instability is influenced by the  weak  and strong   degeneracy of electrons in relativistic plasmas \cite{chanturia2017}. In an another work, it has been emphasized that not only the Raman and Brillouin scattering instabilities are possible, there can be the onset of modulational instabilities of the circularly polarized EMWs at nanoscales in dense quantum plasmas \cite{shukla2012}.
\par
In this work, we present a theoretical study on the stimulated scattering instabilities of intense hf EMWs in a relativistic plasma with degenerate electrons and adiabatic thermal ions. Starting from the EMW equation coupled to the driven (by the EMW's ponderomotive force) equations for low-frequency electron and ion plasma oscillations, we obtain   nonlinear dispersion relations that reveal stimulated Raman scattering (SRS), stimulated Brillouin scattering (SBS) and modulational instabilities (MIs) of EMWs. It is shown that in the field of strong EM radiation, the instabilities develop and the corresponding growth rate  for Raman scattering becomes higher in weakly relativistic degenerate plasmas, however, it can be lower for the Brillouin scattering instability.    
\section{Theoretical Formulation}\label{sec-model}
We consider the nonlinear interaction  of intense  EMWs and the   relativistic plasma composed of degenerate electrons and adiabatic thermal ions. 
%%%%%%%%%%%%%%%%%%%%%%%%%%%%%%%%%%%
Our starting point is the Amp{\'e}re-Maxwell equation 
\begin{equation}
\nabla\times{\bf B}=\frac{1}{c}\left(4\pi{\bf J}+\frac{\partial{\bf E}}{\partial t}\right). \label{amp-maxwell}
\end{equation}
Defining the vector $({\bf A})$ and the scalar $(\phi)$ potentials  by
\begin{equation}
{\bf B}=\nabla\times{\bf A}; ~~{\bf E}=-\frac{1}{c}\frac{\partial{\bf A}}{\partial t}-\nabla\phi, \label{vector-A-eq}
\end{equation}
  and using the Coulomb's gauge condition $\nabla\cdot{\bf A}=0$, we obtain from Eq. \eqref{amp-maxwell} the following EMW equation
  \begin{equation}
  \frac{\partial^2{\bf A}}{\partial t^2}-c^2\nabla^2{\bf A}+c\frac{\partial}{\partial t}(\nabla \phi)-4\pi c{\bf J}=0, \label{em-wave-eq}
  \end{equation}
where ${\bf J}=\sum_{j=e,i} q_j\gamma_jn_j{\bf v}_j$ is the current density with $q_j$, $n_j$ and ${\bf v}_j$ denoting, respectively, the particle's charge, the plasma density in rest frame and the velocity, $c$ is the speed of light in vacuum, and $\gamma_j=1/\sqrt{1-v_j^2/c^2}$  is the Lorentz factor for $j$-th species particle ($j=e,~i$ for electrons and ions).
\par
The dynamics of relativistic electron and ion fluids are given by the following system of equations in which we include the weakly relativistic effects on the particle  motion in the  EMW fields, but fully relativistic effects on the particle thermal motions. Thus, the basic equations read \cite{gomberoff1997,gratton1997}
\begin{equation}
 \begin{split}
 \frac{\gamma_jH_j}{c^2}\frac{d}{dt} (\gamma_j {\bf v}_j)&=q_jn_j\gamma_j \left({\bf E}+\frac{1}{c}{\bf v}_j\times {\bf B} \right)\\
 &-\left(\nabla+\frac{\gamma_j^2{\bf v}_j}{c^2}\frac{d}{d t}\right)P_j, \label{moment-eq}
 \end{split}
 \end{equation}
 \begin{equation}
 \frac{\partial \left(\gamma_jn_j\right)}{\partial t} +\nabla\cdot (\gamma_jn_j\textbf{v}_j)=0,\label{continuity-eq}
 \end{equation}
 \begin{equation}
 \nabla^2 \phi=4\pi e (\gamma_en_e-\gamma_in_i), \label{poisson-eq}
 \end{equation}
 where $d/dt$ stands for $\partial/\partial t+ {\bf v}_j \cdot \nabla$, $e$ is the elementary charge, and $H_j={\cal E}_j+P_j$ is the enthalpy per unit volume measured in the rest frame of each element of the fluid. Here, $P_j$ is the pressure and ${\cal E}_j$ is the total energy density, i.e., ${\cal E}_j=m_jn_jc^2+\bar{\epsilon}_j$ with $\bar{\epsilon}_j$  denoting the internal energy of the fluid and $m_j$ the proper   mass of the $j$-th species particle.
\par 
For simplicity, we assume that the finite amplitude linearly polarized EMWs propagate in the $z$-direction, i.e., all the dynamical variables  vary with $z$ and $t$. So, the transverse electric and magnetic fields are $E_x\equiv E_x(z,t)$, $B_y\equiv B_y(z,t)$ for which ${\bf A}=\left(A_x(z,t),0,0\right)$ and ${\bf v}_j=\left(v_{jx}(z,t),0,v_{jz}(z,t)\right)$, where $v_{jz}$ is generated by the $v_{jx}B_y$ term of the Lorentz force. The longitudinal electric field is given by $E_z(z,t)=-\partial\phi(z,t)/\partial z$ and the longitudinal motion $v_{jz}(z,t)$, coupled with the EMW is   associated with the density variation $n_j(z,t)$.
\par  
Taking the transverse or $x$- component of Eq. \eqref{moment-eq} and using the relations of Eq. \eqref{vector-A-eq} together with the condition $\nabla\cdot{\bf A}=0$, we obtain 
\begin{equation}
   \frac{H_j}{c}\frac{d}{dt} (\gamma_j {v}_{jx})=-\frac{\gamma_jv_{jx}}{c}\frac{dP_j}{dt}-q_jn_j\frac{dA_x}{dt}. \label{moment-eq-x-comp}
  \end{equation}
Next, eliminating $dP_j/dt$ from Eq. \eqref{moment-eq-x-comp} using the relation for adiabatic motion of charged particle, given by,
\begin{equation}
\frac{1}{n_j}\frac{dP_j}{dt}=\frac{d}{dt}\left(\frac{H_j}{n_j}\right), \label{press-adia}
\end{equation}
and integrating the resulting equation with respect to $t$ (assuming  $v_{jx}=0$ for $A_x=0$) we obtain the following transverse (quiver) velocity components of   electrons and ions.
\begin{equation}
v_{jx}=-\frac{q_jn_jc}{\gamma_jH_j}A_x. \label{velo-transv}
\end{equation}
This expression  of  ${ v}_{jx}$ can be substituted in    the Lorentz factor $\gamma_j$ to rewrite it as 
\begin{equation}
 \gamma_j=\sqrt{\frac{1+ q^2_jn_j^2  A^2_x/H^2_j}{1-v_{jz}^2/c^2}}. \label{gamma}
\end{equation} 
On the other hand, considering the parallel or $z$- component of Eq. \eqref{moment-eq}, and eliminating $dP_j/dt$ and $v_{jx}$   by   Eqs.  \eqref{press-adia} and \eqref{velo-transv}, we obtain 
\begin{equation}
\frac{d}{dt}\left(\frac{\gamma_jH_jv_{jz}}{n_jc^2}\right)=-q_j\left(\frac{\partial\phi}{\partial z}+\frac{1}{2}\frac{q_jn_j}{\gamma_jH_j}\frac{\partial A_x^2}{\partial z}\right)-\frac{1}{\gamma_jn_j}\frac{\partial P_j}{\partial z}, \label{moment-eq-parallel}
\end{equation}
Furthermore, using Eq. \eqref{velo-transv}, the transverse component of the EMW equation \eqref{em-wave-eq} can be obtained as
\begin{equation}
\left(\frac{\partial^2}{\partial t^2}- c^2\frac{\partial^2}{\partial z^2}\right) {A}_x=4\pi ec\left(N_iv_{ix}-N_ev_{ex}\right), \label{em-wave-transv}  
\end{equation}
where we denote $N_j\equiv\gamma_jn_j$. The longitudinal component of Eq. \eqref{em-wave-eq} gives
\begin{equation}
\frac{\partial^2\phi}{\partial t\partial z}=4\pi e\left(N_i v_{iz}-N_e v_{ez}\right).\label{em-wave-longi}
\end{equation}
Also, in one-dimension, the equations  \eqref{continuity-eq} and \eqref{poisson-eq} reduce to
\begin{equation}
 \frac{\partial \left(\gamma_jn_j\right)}{\partial t} +\frac{\partial}{\partial z} (\gamma_jn_j {v}_{jz})=0,\label{continuity-eq-reduced}
 \end{equation}
 \begin{equation}
 \frac{\partial^2 \phi}{\partial z^2}=4\pi e (\gamma_en_e-\gamma_in_i), \label{poisson-eq-reduced}
 \end{equation}
Equations \eqref{moment-eq-parallel}, \eqref{continuity-eq-reduced} and \eqref{poisson-eq-reduced} are the required equations for the description of low-frequency electron and ion plasma oscillations that are driven by the EMW's ponderomotive force and coupled to the EMWs given by Eq. \eqref{em-wave-transv}. Here we note that since the electrostatic potential $\phi$, associated with the electric field $E_z$, is created due to the density variations of charged particles we replace  Eq. \eqref{em-wave-longi} by the Poisson equation \eqref{poisson-eq-reduced} for the low-frequency plasma oscillations. In fact, Eq. \eqref{em-wave-longi} will confirm some estimates for the density perturbations and potential, to be considered shortly.    In order to close the system of Eqs. \eqref{moment-eq-parallel}$-$ \eqref{poisson-eq-reduced} we need   expressions for the pressures $P_j$.  
For relativistic degenerate electrons, the pressure $P_e$ and the total energy density ${\cal E}_e$ are   given by  \cite{chandrasekhar1935}  
 \begin{equation}
 \begin{split}
 &\left(P_e,{\cal E}_e\right)=\frac{m_e^4c^5}{3\pi^2\hbar^3}\left[f(R),~R^3\left(1+R^2\right)^{1/2}-f(R)\right],\\
 &f(R)=\frac18\left[ R\left(2R^2-3\right)\left(1+R^2\right)^{1/2}+3\sinh^{-1}R\right],
 \end{split} \label{pressure}
\end{equation}  
where $\hbar=h/2\pi$ is the reduced Planck's constant, $R=p_{Fe}/m_ec=\hbar\left(3\pi^2n_e\right)^{1/3}/m_ec$ is the dimensionless degeneracy parameter, and $H_e\equiv P_e+{\cal E}_e=n_em_ec^2\sqrt{1+R^2}$. 
\par 
For adiabatic thermal ions we may write $P_i=P_{i0}\left(n_i/n_0\right)^{\Gamma}$ with a polytropic index $\Gamma$, given by $4/3\leq\Gamma\leq5/3$   such that   in the classical (low-energy plasma) limit, $\Gamma=5/3$ and $P_i\ll n_im_ic^2$, while in the ultra-relativistic limit we have $\Gamma=4/3$ and $P_i\gg n_im_ic^2$. 
\par 
In what follows, we derive a reduced set of equations from Eqs. \eqref{moment-eq-parallel}$-$ \eqref{poisson-eq-reduced} for the slow motion approximation of relativistic dynamics of electrons and ions, i.e., when the perturbations $\gamma_{j}\approx 1+(1/2)\left({\bf v}_{j}/c\right)^2$. We also assume that $|q_j|n_jA_x/H_j \sim o(\epsilon)$ for which Eq. \eqref{velo-transv} yields $v_{jx}/c\sim o(\epsilon)$, and so from Eqs. \eqref{moment-eq-parallel} and \eqref{em-wave-longi} one can verify that the perturbations  $\phi,~n_{j1},~v_{jz}\sim o(\epsilon^2)$.  
Thus,   Eq. \eqref{gamma} gives
\begin{equation}
\gamma_j^2\approx 1+\left(\frac{q_jn_jA_x}{H_j}\right)^2+o(\epsilon^4), \label{gamma-reduced}
\end{equation}
and,  so, from Eq. \eqref{velo-transv}
\begin{equation}
\frac{v_{jx}}{c}\approx-\frac{q_jn_j}{H_j}A_x\left[1-\frac{1}{2}\left(\frac{q_jn_jA_x}{H_j}\right)^2\right]. \label{velo-trans-reducd}
\end{equation}    
Next, we linearize Eqs. \eqref{moment-eq-parallel}$-$\eqref{poisson-eq-reduced} and \eqref{velo-trans-reducd} about the equilibrium state, i.e., assuming $n_j=n_0+n_{j1},~H_j=H_{j0}+H_{j1}$ and $N_j\equiv \gamma_jn_j=n_0+N_{j1}$  etc., and following  Ref. \onlinecite{gratton1997} we obtain from Eq. \eqref{velo-trans-reducd}
\begin{equation}
\begin{split}
\frac{v_{jx}}{c} \approx &-\frac{q_jn_0}{H_{j0}}A_x\left[1-\frac{1}{2}\left(\frac{q_jn_0A_x}{H_{j0}}\right)^2+\frac{n_{j1}}{n_0}-\frac{H_{j1}}{H_{j0}}\right]\\
&+o(\epsilon^4), \label{velo-trans-linear}
\end{split}
\end{equation}
and from Eq. \eqref{moment-eq-parallel}
\begin{equation}
\begin{split}
\frac{1}{c}\frac{\partial}{\partial t}\left(\frac{v_{jz}}{c}\right)=&-\frac{q_jn_0}{H_{j0}}\frac{\partial\phi}{\partial z}-\frac{1}{2}\left(\frac{q_jn_0}{H_{j0}}\right)^2\frac{\partial A_x^2}{\partial z}\\
&-\frac{1}{{H_j0}}\left(\frac{\partial P_j}{\partial n_j}\right)_{n_j=n_0}\frac{\partial n_{j1}}{\partial z}+o(\epsilon^4). \label{moment-eq-parallel-reducd}
\end{split}
\end{equation}
For relativistic degenerate electrons we have  $\left({\partial P_e}/{\partial n_e}\right)_{n_e=n_0}=2{\cal E}_F/3\sqrt{1+R_0^2}$ where ${\cal E}_F=\hbar^2(3\pi^2n_0)^{2/3}/2m_e$ is the electron Fermi energy,  $R_0=p_0/m_ec$ is the dimensionless  parameter which measures the strength of the plasma degeneracy and $p_0=(3h^3n_0/8\pi)^{1/3}$  is  the momentum of electrons on the Fermi surface. On the other hand, for adiabatic thermal ions, $\left({\partial P_i}/{\partial n_i}\right)_{n_i=n_0}=P_{i0}\Gamma/n_0$.
\par
  First, we consider the  electrostatic electron plasma oscillations that are driven by the EMW's poderomotive force on the time scale of electron plasma period $\omega_p^{-1}\equiv 1/\sqrt{4\pi n_0e^2/m_e}$. In this case, ions can be considered as immobile for which the ion density perturbation is zero. Thus,  we eliminate $v_{ez}$ from Eq. \eqref{moment-eq-parallel-reducd} and the linearized form of Eq. \eqref{continuity-eq-reduced}, i.e., $\partial_t\left(N_{j1}/n_0\right)=-\partial_zv_{jz}$, and finally eliminate the perturbed potential $\phi_1$   using the Poisson equation $\partial_z^2\phi=4\pi eN_{e1}$ to obtain the following equation  for the  density variation of  low-frequency electron plasma oscillations that are reinforced by the EMW's ponderomotive force.
   \begin{equation}
\left(\frac{\partial^2}{\partial t^2}-\delta_e\frac{\partial^2}{\partial z^2}+1\right)N=\frac{1}{2} (1-\delta_e)\frac{\partial^2 A^2}{\partial z^2}. \label{electron-main-eq}
\end{equation}
Substituting the expression of $v_{jx}$ from Eq. \eqref{velo-trans-linear} into   Eq. \eqref{em-wave-transv}, and neglecting the terms  $\sim o(\epsilon^4)$, we obtain  the EMW equation
\begin{equation}
\left(\frac{\partial^2}{\partial t^2}- \frac{\partial^2}{\partial z^2}+1\right)A + (1-\delta_e)N  A=0.\label{cpem-main-eq}
\end{equation} 
In Eqs. \eqref{electron-main-eq} and \eqref{cpem-main-eq},  we have considered the normalizations as $t\rightarrow t\sqrt{\eta_e}\omega_p $, $z\rightarrow z\sqrt{\eta_e}\omega_p/c$, $A\equiv \eta_e eA_x/m_ec^2$, $N\equiv N_{e1}/n_0\ll1$ with $\eta_e=1/\sqrt{1+R_0^2}$ and $\delta_e=(1-\eta_e^2)/3$.     
The case with $R_0\ll1~(\gg1)$ corresponds to the weakly relativistic (ultra-relativistic) degenerate plasmas.  
\par
Second, we consider the driven  low-frequency ion plasma oscillations.  By the similar manner as for Eq. \eqref{electron-main-eq},   and assuming the  quasineutrality $N_{i1}\approx N_{e1}$ (valid for long-wavelength low-frequency perturbations),   we obtain the following   wave equation for  ion plasma oscillations.
\begin{equation}
\left(\frac{\partial^2}{\partial t^2}-\sigma \frac{\partial^2}{\partial z^2}\right)N_e=\frac{1}{2}\beta \frac{\partial^2 A^2}{\partial z^2},\label{ion-main-eq}
\end{equation}
where $\sigma=\left(\eta_e\delta_i+\eta_i\delta_e\right)/\left(\eta_e+\eta_i\right) $, $\beta=\left(\eta_i/\eta_e\right)\left[\eta_e(1-\delta_e)+\eta_i(1-\delta_i) \right]/ \left(\eta_e+\eta_i\right)$,   $\delta_i=P_{i0}\Gamma/H_{i0}$, and $\eta_i=\left(m_e/m_i\right)\left(1-3\delta_i/2\right)$.  
Equations \eqref{electron-main-eq} $-$ \eqref{ion-main-eq} are the desired  equations for studying, e.g., the localization of EM solitons \cite{sundar2016,gratton1997}, the generation of wakefields \cite{holkundkar2018,misra2010},    the onset of stimulated scattering instabilities \cite{chanturia2017,shukla2012}  in relativistic plasmas with degenerate electrons and adiabatic thermal ions.
\section{Derivation and analysis of nonlinear dispersion relation}\label{sec-disp-relation}
To investigate the characteristics of SRS,  SBS and  MIs of a constant amplitude pump that is scattered off   electron and ion plasma modes, and a spectrum of nonresonant electron and ion density perturbations,  we express the   potential $A$ as 
\begin{equation}
\begin{split}
A =&A_0 \exp(ik_0 x-i\omega_0 t)+\text{c.c.}\\
&+\sum_{+,-}{A}_{\pm} \exp(ik_\pm x-i\omega_\pm t), \label{A-perp-eq}
\end{split}
\end{equation}
where \textit{c.c.} denotes the complex conjugate. The subscripts $0$ and $\pm$ stand for the EMW pump and EMW sidebands, respectively, and $\omega_{\pm}=\Omega \pm \omega_0$, $k_\pm=k\pm k_0$ are the frequencies and wave numbers of the EMW sidebands  that are generated  due to the interactions of the pump wave $(\omega_0, k_0)$ with preexisted low-frequency  electrostatic plasma oscillations $(\Omega,K)$. Substituting Eq. \eqref{A-perp-eq} into the coupled sets of equations [\eqref{electron-main-eq}, \eqref{cpem-main-eq}] and [\eqref{cpem-main-eq}, \eqref{ion-main-eq}], and assuming that $N\sim\exp(iKz-i\Omega t)$ we, respectively,  obtain the nonlinear dispersion relations for SRS and SBS of EMWs:
\begin{equation}
S_R=(1-\delta_e)K^2\sum_{+,-} \frac{1}{D_{\pm}}\big{|{A}_0|}^2,\label{s-r-eq}
\end{equation}
and 
\begin{equation}
S_B=\beta K^2\sum_{+,-} \frac{1}{D_{\pm}}\big{|{A}_0|}^2,\label{s-b-eq}
\end{equation}
where $S_R\equiv \Omega^2-\delta_eK^2-1$, $S_B\equiv \Omega^2-\sigma K^2$, $D_\pm=\omega_\pm^2-k_\pm^2-1\approx \pm 2\omega_0(\Omega-Kv_g\mp\delta)$ with $v_g=k_0/\omega_0$ denoting the dimensionless group velocity of the EMW pump, $\omega_0=\sqrt{1+k_0^2}$ is the pump frequency, and $\delta=K^2/2\omega_0$ is the dimensionless nonlinear frequency shift. The dispersion relations \eqref{s-r-eq} and \eqref{s-b-eq} have the forms similar to those obtained in Ref. \onlinecite{shukla2012}. In the latter, the authors investigated the stimulated scattering instabilities in nonrelativistic quantum plasmas with the effects of particle dispersion and exchange-correlation. However, we have considered the relativistic fluid model in a self-consistent manner quite distinctive from the model in Ref. \onlinecite{shukla2012}.  The solutions of Eqs. \eqref{s-r-eq} and \eqref{s-b-eq}, in fact, represent forward and backward SRS and SBS.
\par
In the absence of the pump,   $S_R=S_B=0$, and  we have the following dispersion relations for electron (Langmuir) and ion plasma oscillations.
\begin{equation}
\Omega_L=\sqrt{1+\delta_eK^2},~~\Omega_I=\sqrt{\sigma}K, \label{Omega-eq}
\end{equation}
\par
Next, we obtain the growth rates   for SRS and SBS instabilities, as well as for the MI of a pump of constant amplitude that is scattered off electron and ion plasma waves. For three-wave decay interactions, the maximum growth rates can be obtained  when the scattered wave is also resonant, i.e.,  $D_-=0~(\Omega=\Omega_L)$ which gives
\begin{equation}
\omega_0=\left(1+\delta_e K^2\right)^{1/2}+\left[1+(K-k_0)^2\right]^{1/2}. \label{omega0}
\end{equation}
%%%%%%%%%%%%%%%%%%%%%%%%%%%%%%%%%%%%%%%%%%%%%%%%%%
\begin{figure*}
\includegraphics[scale=0.5]{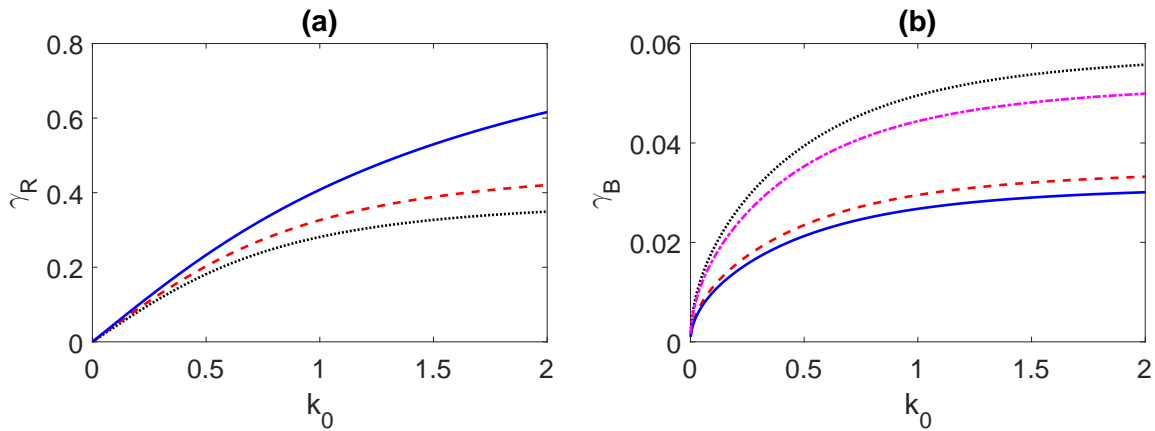}
\caption{The growth rates, given by Eq. \eqref{gamma-eq},  are shown for the   stimulated Raman scattering $(\gamma_R)$ and the stimulated Brillouin scattering $(\gamma_B)$  instabilities. In the subplots (a) and (b), the solid, dashed and dotted lines   correspond  to the regimes with number densities $n_0=10^{28}$ cm$^{-3}$, $10^{30}$ cm$^{-3}$ and    $10^{32}$ cm$^{-3}$ for which $R_0=0.257,~1.19$,   and $5.54$ respectively, and with a fixed value of $A_0=0.5$. In the subplot (b),  $T_i=3\times10^{10}$ K and $\Gamma=5/3$    for the solid, dashed and dotted lines, while for the    dash-dotted line,    $T_i=5\times10^{10}$ K and $\Gamma=5/3$   such that $P_{i0}<<m_in_0c^2$ (low-energy ions) holds. The other fixed values for the dash-dotted line are    $n_0=10^{32}$ cm$^{-3}$ and $A_0=0.5$.   }
\label{fig1}
\end{figure*}
%%%%%%%%%%%%%%%%%%%%%%%%%%%%%%%%%%%%%%%%%%%%%%%%%%
%%%%%%%%%%%%%%%%%%%%
\begin{figure*}
\includegraphics[scale=0.5]{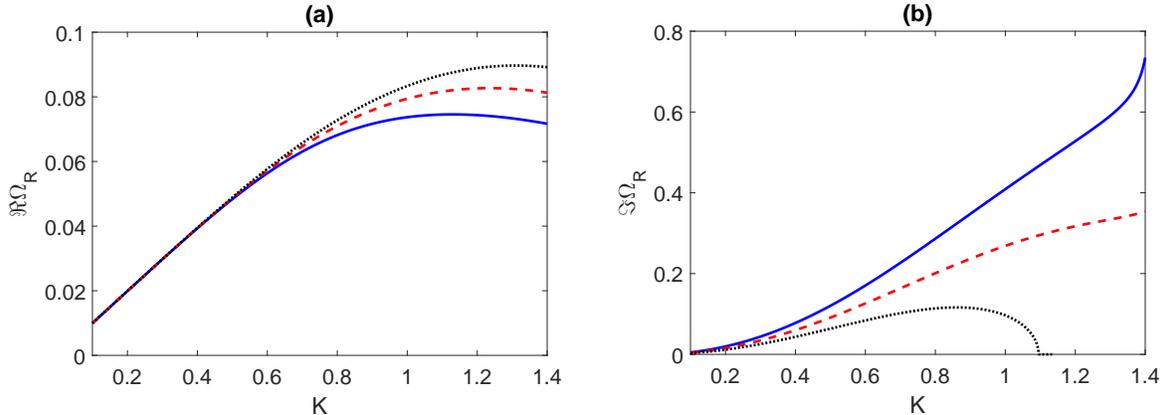}
\caption{The frequency shift $(\Re \Omega)$ and the growth rate of instability $(\Im \Omega)$ [numerical solutions of  Eq. \eqref{modified-s-r-eq}] for the modulational instability of EMWs that are scattered off electron density perturbations are shown with respect to the wave number of modulation $K$ for the same set of parameters as for Fig. \ref{fig1} (a). It is seen that higher the number density, lower is the growth rate with a cut-off at  a lower value of $K$. }
\label{fig2}
\end{figure*}
%%%%%%%%%%%%%%%%%%%%%%%%%%%%%%%%%%%%%%%%%%%%%%%%%%
 %%%%%%%%%%%%%%%%%%%%%%%%%%%%%%%%%%%%%%%%%%%%%%%%%%
\begin{figure*}
\includegraphics[scale=0.5]{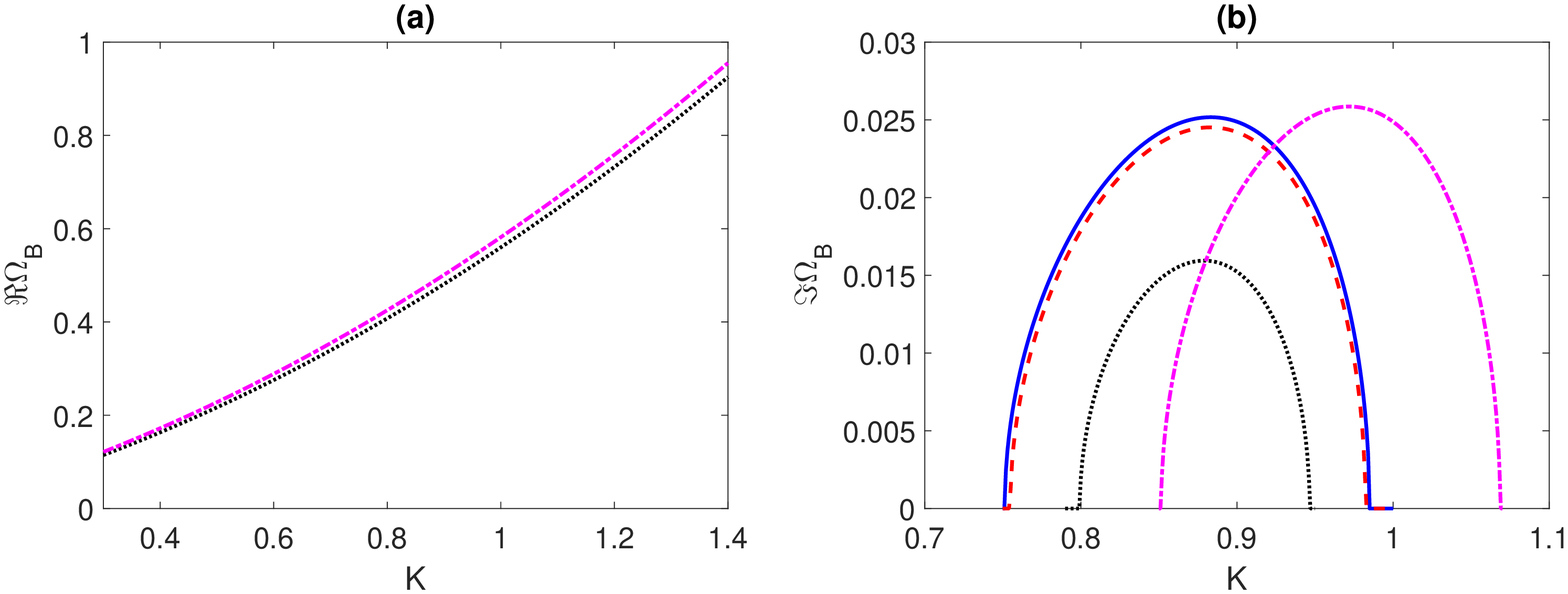}
\caption{The frequency shift $(\Re \Omega)$ and the growth rate of instability  $(\Im \Omega)$ [numerical solutions of  Eq. \eqref{modified-s-b-eq}] for the modulational instability of EMWs that are scattered off ion density perturbations  are shown with respect to the wave number of modulation $K$.  The solid, dashed and dotted lines   correspond  to the regimes with number densities $n_0=10^{28}$ cm$^{-3}$, $10^{30}$ cm$^{-3}$ and    $10^{32}$ cm$^{-3}$   with a fixed value of $A_0\sim0.5$ and $T_i=3\times10^{12}$ K, $\Gamma=5/3$ such that $P_{i0}< m_in_0c^2$ (low energy ions). The    dash-dotted line is for a different ion temperature   $T_i=4\times10^{12}$ K, $\Gamma=5/3$ such that $P_{i0}<m_in_0c^2$   with a fixed $n_0=10^{32}$ cm$^{-3}$  and $A_0\sim0.5$.}
\label{fig3}
\end{figure*}
%%%%%%%%%%%%%%%%%%%%%%%%%%%%%%%%%%%%%%%%%%%%%%%%%%
\par
Thus, in an underdense plasma with $\omega_0>2$, one can show using Eq. \eqref{omega0} that the wave number $K$ of resonant modes lies in $0<K<2k_0$. The modes with $K<k_0$ and $K>k_0$, respectively,  correspond to forward and backward SRS/SBS instabilities. Neglecting the nonresonant terms $(\sim 1/D_+)$ from Eqs. \eqref{s-r-eq} and \eqref{s-b-eq}, and letting $\Omega=(Kv_g-\delta)+i\gamma_{R,B}= \Omega_{R,B}+i\gamma_{R,B}$,  we obtain the growth rates   for the Raman and Brillouin backscattering $(K\simeq2k_0>k_0)$ instabilities: 
\begin{equation}
\gamma_R=\frac{k_0\sqrt{1-\delta_e}}{\sqrt{\omega_0\Omega_R}}|A_0|, ~~
\gamma_B=\frac{k_0\sqrt{\beta}}{\sqrt{\omega_0\Omega_B}} |A_0|, \label{gamma-eq}
\end{equation}
where $\Omega_{R,B}=\Omega_{L,I}$ at $K\simeq2k_0$, i.e., $\Omega_R=\sqrt{1+4\delta_e k_0^2}$ and $\Omega_B=2k_0\sqrt{\sigma}$.
The explicit dependencies of  $\gamma_R$ and $\gamma_B$ on $\delta_e$ and $\beta$ show that  the growth rates are significantly modified by the  relativistic degenerate pressure  of electrons and the adiabatic thermal pressure   of ions.  
\par
Next, for the modulational instabilities of EMWs associated with the nonresonant electron and ion density perturbations, we retain both $D_\pm(\neq0)$ and $S_{R,B}(\neq 0)$ in Eqs. \eqref{s-r-eq} and \eqref{s-b-eq}. Thus, we obtain \cite{shukla2012}
\begin{equation}
S_R\left[\left(\Omega-Kv_g\right)^2-\delta^2\right]=\frac{\delta(1-\delta_e) }{\omega_0}K^2|A_0|^2, \label{modified-s-r-eq}
\end{equation}
and 
\begin{equation}
S_B\left[\left(\Omega-Kv_g\right)^2-\delta^2\right]=\frac{\beta\delta }{\omega_0}K^2|A_0|^2.\label{modified-s-b-eq}
\end{equation}
Equations \eqref{modified-s-r-eq} and \eqref{modified-s-b-eq} can be solved numerically to ascertain the growth rates of MIs of EMWs that are scattered by the nonresonant electron and ion density perturbations, which we will perform in Sec. \ref{sec-results-discus}.   
\section{Results and discussion}\label{sec-results-discus}
We  numerically investigate the characteristics of the growth rates for SRS and SBS instabilities given by Eq. \eqref{gamma-eq}. The results are displayed in Fig. \ref{fig1}. From the subplot \ref{fig1} (a), it is seen that as one goes from the regimes of weak  to strong relativistic degenerate plasmas (by increasing the number density and so are the values of both $R_0$ and $\delta_e$), the growth rate for SRS instability $\gamma_R$ is reduced (see the solid, dashed and dotted lines).  It follows  that even in the field of strong EMW radiation, the growth rate of instability is enhanced in the regimes of weakly relativistic degenerate plasmas. On the other hand, subplot \ref{fig1} (b) shows that the influence of the degenerate pressure of electrons is also pronounced due to the    dependency of $\beta$ on the degeneracy parameter $R_0$.  In this case, the growth rate is significantly enhanced in highly dense plasmas with strong degeneracy of electrons, i.e., $R_0>1$.    However, as the ion thermal energy increases, the growth rate of SBS instability is found to be reduced (see the dotted and dash-dotted lines). 
\par
Next, we numerically solve Eqs. \eqref{modified-s-r-eq} and \eqref{modified-s-b-eq} to obtain the frequency shifts $(\Re\Omega)$ and the growth rates  $(\Im\Omega)$  of instabilities of  the EMW  envelope of constant amplitude    that is scattered off and modulated by  the nonresonant electron and ion density perturbations. The corresponding results are exhibited in Figs. \ref{fig2} and \ref{fig3} respectively.  From Fig. \ref{fig2} (a),  we notice that the frequency of modulation  is always up shifted and it increases as one goes from weakly relativistic to strong or ultra-relativistic regimes of degenerate electrons.   On the other hand, subplot  Fig. \ref{fig2} (b) shows that higher the concentration of number density (strong relativistic degeneracy), lower is the growth rate of instability with cut-offs at lower wave numbers of modulation $K$ (see the  dotted line).   Figure \ref{fig3} (a) shows  that   the influence of the degenerate pressure of electrons on the frequency shift is not so pronounced, however, it increases with increasing values of the ion thermal energy.  On the other hand, both the    thermal pressure of ions and the degenerate pressure of electrons significantly modify the   growth rate of instability of EMWs  under the modulation of ion density perturbations.  We find that  the value of $(\Im\Omega_B)$ is reduced with an increase of the particle number density (see the solid and dashed lines) having a cut-off at a lower value  of $K$. Such a reduction is significant in the high-density regimes (see the dotted line) or in the ultra-relativistic regimes of degenerate electrons.  In contrast, an enhancement of the growth rate is  seen  by  increasing   the ion thermal energy with a cut-off at higher value of $K$.
 \section{Conclusion} \label{sec-conclu}
 We have investigated the nonlinear interactions of finite amplitude high-frequency EMWs with low-frequency electrostatic electron and ion density perturbations that are driven by the EMW's ponderomotive force in an unmagnetized relativistic plasma with degenerate electrons and thermal ions. At the time scale of electron plasma period when ions do not respond, it is shown that the Langmuir wave spectra is significantly modified by the relativistic degenerate electrons, and are excited by the EMWs due to stimulated Raman scattering instability. In this case,    the growth rate of instability is shown to be reduced   in strong relativistic degenerate plasmas.    Furthermore, the inclusion of ion dynamics provides also the possibility of  low-frequency ion-acoustic waves that are modified by the degenerate pressure of  electrons and the ion thermal pressure, and are excited by  the EMW due to   Brillouin scattering instability. In this case, the instability growth rate is seen to be significantly enhanced in the regime of strong degeneracy of electrons and  low ion thermal energy.     We have also shown the possibility of   the modulational instability of EMWs due to nonresonant electron and ion density perturbations. The characteristics of the frequency shifts and the growth rates of modulational instability are found to be quite distinctive from those in nonrelativistic regimes \cite{shukla2012}. 
 \par
 To conclude, the results of stimulated  scattering instabilities, as well as the modulational instability of  intense EMWs in a relativistic degenerate plasma is highly pertinent to understanding the salient features of enhanced density fluctuations and the dynamics of $X$-ray pulses that may emanate from compact astrophysical objects. The results can also be useful in the next-generation highly intense laser produced solid density compressed plasma experiments.    
%%%%%%%%%%%%%%%%%%%%%%%%%%%%%%%
 \acknowledgments{This work was supported by UGC-SAP (DRS, Phase III) with 
Sanction  order No.  F.510/3/DRS-III/2015(SAPI),  and UGC-MRP with F. No.
43-539/2014(SR) and FD Diary No. 3668.}
%%%%%%%%%%%%%%%%%%%%%%%%%%%%%%%5

\end{document}